\documentclass[12pt]{article}
\usepackage{amssymb,latexsym,amsmath}

\usepackage[activeacute,english]{babel}
\usepackage{fancyhdr}
\usepackage{textcomp}
\usepackage{amsfonts}
\usepackage{graphicx}
\usepackage{graphics}
\usepackage[colorlinks=true]{hyperref}
\usepackage[active]{srcltx}

\newcommand{\be}{\begin{equation}}
\newcommand{\ee}{\end{equation}}
\newcommand{\ba}{\begin{array}}
\newcommand{\ea}{\end{array}}
\newcommand{\bea}{\begin{eqnarray}}
\newcommand{\eea}{\end{eqnarray}}

\def\C{{\mathbb C}}
\newcommand{\bi}{\begin{itemize}}
\newcommand{\ei}{\end{itemize}}

\newtheorem{theorem}{Theorem}
\newtheorem{definition}{Definition}

\newtheorem{example}{Example}

\begin{document}
\title{Toward a classification of semidegenerate 3D superintegrable systems}
\author{{\bf M.A. Escobar-Ruiz}\\
{\sl Instituto de Ciencias Nucleares, UNAM}\\
{ \sl Apartado Postal 70-543, 04510 Mexico D.F. MEXICO}\\
{\sl and School of Mathematics, University of Minnesota}\\
{\sl mauricio.escobar@nucleares.unam.mx}\\
{\bf  and  Willard Miller, Jr.}\\
{\sl School of Mathematics, University of Minnesota,}\\
{\sl Minneapolis, Minnesota,
55455, U.S.A.}\\
{\sl miller@ima.umn.edu}}
\date{\today}
\maketitle
\abstract{Superintegrable systems of 2nd order in 3 dimensions with exactly 3-parameter potentials are intriguing objects. Next to the nondegenerate 4-parameter potential systems
they admit the maximum number of symmetry operators but their symmetry algebras don't close and not enough is known about their structure to give a
complete classification. Some examples are known for which the 3-parameter system can be extended to a 4th order superintegrable system with 
a 4-parameter potential and 6 linearly independent symmetry generators. In this paper we use B\^ocher contractions of the conformal Lie algebra $so(5,\C)$ to itself
to generate a large family of 3-parameter systems with 4th order extensions, on a variety of manifolds, and all from B\^ocher contractions of a single ``generic'' system 
on the 3-sphere. We give a contraction scheme relating these systems. The results have myriad applications for finding explicit solutions for both quantum and classical systems.}

\section{Introduction}
Superintegrable quantum mechanical systems admit the maximum possible symmetry and this forces analytic and algebraic solvability. These systems appear in a wide variety of modern physical and mathematical theories, 
from semiconductors to supersymmetric field theories, \cite{SCQS,MPW2013}. Superintegrable systems of 2nd order are of particular interest due primarily to their
connection with separation of variables. The special functions of mathematical physics and 
their properties are closely related to their origin and use in providing explicit solutions for 2nd order superintegrable systems. 
The structure theory for 2D 2nd order superintegrable  Helmholtz and Laplace equations has been worked out in its entirety, 
\cite{KKM20041,Laplace2011, CKP2015,KMS2016,KMS2016a}. 
There is a single family of 
superintegrable systems with generating symmetry operators that are functionally 
linearly dependent;  the remaining (functionally linearly independent) systems are nondegenerate (3 parameter Helmholtz potentials) and degenerate (1 parameter potentials). 
Every functionally linearly independent system is obtainable from the generic system on the 2-sphere through a sequence of restrictions, 
B\^ocher contractions and St\"ackel transforms. Each Laplace equation is a
St\"ackel equivalence class of Helmholtz systems and always contains a constant curvature space representative.
The nondegenerate systems
always have 3 functionally independent 2nd order generators which determine a quadratic algebra that closes at order 6.

However, the   hierarchy of 3D 2nd order superintegrable Helmholtz and Laplace equations is  only partially worked out. 
There are now multiple functionally linearly dependent systems 
(such as the Calogero 3-body system on the line) and we are not aware of a classification for them. 
All of the nondegenerate (4-parameter Helmholtz potential) systems are known, \cite{CapelKress}. These
have 5 functionally linearly independent, contained in   6 linearly  independent (but functionally dependent) 2nd order generators 
which determine a quadratic algebra that closes at order 6. The 
functional dependence is described by a relation at order 8. Every Laplace equation is again a
St\"ackel equivalence class of Helmholtz systems and always contains a constant curvature space representative. Every nondegenerate
system is obtainable from the generic system on the 3-sphere through a sequence of B\^ocher contractions and St\"ackel transforms. 

Immediately below the 4-parameter Helmholtz systems in the 3D hierarchy are the 3-parameter systems. These admit 5 functionally linearly
independent 2nd order generators. The first recognition of the  special significance   of 3D Helmholtz superintegrable systems 
that had only 3-parameter potentials
was in the paper \cite{EVA} by Evans.  The most important early example studied was the {\bf extended Coulomb system}. 
The  Schr\"odinger operator in Cartesian coordinates $(x,y,z)$ can be written as
\be
\label{1.66}
H_{cI}=\partial^2_{x}+\partial^2_{y}+\partial ^2_{z}-
\frac{a }{ \sqrt{x^2+y^2+ z^2}} +
\frac{a_1}{ x^2}+
\frac{a_2}{ y^2}.
\ee

It admits symmetries (here $J_{12 }=x\partial_{y}-y\partial_{x},J_{13 }=x\partial_{z}-z\partial_{x},J_{23 }=y\partial_{z}-z\partial_{y}$),
\be
\label{1.67}
L_{12}=J^2_{12}+a_2\frac{x^2}{ y^2} +
a_1\frac{y^2}{ x^2},\quad 
L_{13}=J^2_{13}+a_1\frac{z^2}{ x^2},\ L_{23}=J^2_{23}+a_2\frac{z^2}{ y^2},
\ee
\be
L=-\frac12\left(\{\partial_{x}, J_{13}\}+\{\partial_{y}, J_{23}\}
+2a_1\frac{z}{ x^2}+2a_2\frac{z}{ y^2}\right) -
\frac{a\, z}{ 2\sqrt{x^2+y^2 + z^2}}.
\label{1.69}
\ee
In this case the symmetry algebra doesn't close under commutation, \cite{KKM2007}.
However, Verrier and Evans, \cite{VE2006}, showed that this system could be extended to a 4-parameter potential 
corresponding to a 4th order superintegrable system with 5 generators: four 2nd order and one 4th order. 
Later it was shown that this extended system admitted
a second independent 4th order generator and that the 6 linearly independent symmetries determined an algebra that closed at order 10, 
while the
functional identity relating the 6 generators was order 12, \cite{TD2011,KKM2013}. The extended system and its generating symmetries are as follows:
\be\label{Hc} H_{cII}=\partial_{x}^2+\partial_{y}^2+\partial_{z}^2+\frac{a}{\sqrt{x^2+y^2+z^2}}+\frac{a_1}{x^2}+\frac{a_2}{y^2}+\frac{a_3}{z^2},\ee
\[ I_{12}= J_{12}^2+\frac{a_1y^2}{x^2}+\frac{a_2x^2}{y^2},\  I_{13}= J_{13}^2+\frac{a_1z^2}{x^2}+\frac{a_3x^2}{z^2},\  I_{23}=  J_{23}^2
+\frac{a_2z^2}{y^2}+\frac{a_3y^2}{z^2},\]
\[ M_3=\frac12( \{J_{23},\partial_{y}\})-\{J_{31},\partial_{x}\})-z(\frac{a}{2\sqrt{x^2+y^2+z^2}}+\frac{a_1}{x^2}+\frac{a_2}{y^2}+\frac{a_3}{z^2}),\]
\[ M_1=\frac12( \{J_{31},\partial_{z}\})-\{J_{12},\partial_{y}\})-x(\frac{a}{2\sqrt{x^2+y^2+z^2}}+\frac{a_2}{y^2}+\frac{a_3}{z^2}
+\frac{a_1}{x^2}),\]
\[ J_0:=\frac14\left(-16M_3^2-\{(\{x,\partial_{x}\}+\{y,\partial_{y}\}+\{z,\partial_{z}\})^2,{2a_3}{z}\}\right),\]
\[ {J'}_0:=\frac14\left(-16M_1^2-\{(\{x,\partial_{x}\}+\{y,\partial_{y}\}+\{z,\partial_{z}\})^2,{2a_1}{x^2}\}\right) .\]
Here, $\{A,B\}=AB+BA$, the operator symmetrizer. A basis of generators for the symmetry operators is $\{ H,I_{12},I_{13},I_{23}, J_0,{J'}_0\}$.

Another example of the extension of a 3-parameter potential  2nd order superintegrable system to a 4-parameter potential 4th order superintegrable system
is the extended anisotropic oscillator, \cite{RTW2008,EV2008}.  
Several other examples of 3-parameter 2nd order superintegrable systems have been reported and there are 
some structure results, \cite{KKM20041}. In particular, every 3-parameter system is multiseparable and St\"ackel equivalent to a 3-parameter 
constant curvature space system. A 3-parameter system can be extended to a nondegenerate 4-parameter system if and only if it 
admits 6 linearly independent 2nd order symmetries. 
The 5 generators of the symmetry algebra of a truly 3-parameter system do not determine a closed structure in the usual manner. We call such a truly 3-parameter 
system {\it semidegenerate}:
\begin{definition} A 3D Helmholtz superintegrable system on a conformally flat space is {\bf semidegenerate} provided it satisfies the following conditions \cite{KKM20041}:
\begin{enumerate} \item It is 2nd order superintegrable, i.e., it admits 5 functionally independent 2nd order symmetries.
\item It admits a 3-parameter potential $V({\bf x})=a_1V^{(1)}({\bf x}) +a_2V^{(2)}({\bf x}) +a_3V^{(3)}({\bf x})$ where the set $\{ V^{(1)},V^{(2)},V^{(3)}\}$ is functionally independent.
\item It fails to be nondegenerate, i.e., it does not admit 6 functionally independent 2nd order symmetries.
\end{enumerate}
\end{definition}
There are analogous definitions for semidegenerate Laplace and classical systems. 
In the hierarchy of 3D Helmholtz and Laplace superintegrable systems the semidegenerate systems are just one step below the 
nondegenerate (4-parameter) systems.

Up to now there has been no regular procedure for deriving 
semidegenerate systems and determining which can be extended to 4-parameter systems of 4th order. We provide a partial solution to this problem.
Since our past experience is that 
the most `generic' systems are those on spheres, we first find 
such a 4th order system on the 3-sphere and then use the tools of B\^ocher contractions and St\"ackel transforms to obtain other 
systems as limits.

The new 3-parameter system on the 3-sphere is (in Laplace equation form) $ H_{coulsphere}\Theta=0$ where
\be\label{Hcoulsphere} H_{coulsphere}=\Delta_S+\frac{a_1s_4}{\sqrt{s_1^2+s_2^2+s_3^2}}+\frac{a_2}{s_1^2}+\frac{a_3}{s_2^2}+a_4,\ee
and $\Delta_S=\sum_{1\le j<k\le 4}(s_j\partial_{s_k}-s_k\partial_{s_j})^2$,
is the Laplace-Beltrami operator on the 3-sphere.
A basis of 2nd order symmetries is\break  $\{H_{coulsphere}, L_{12},L_{13},L_{23},L\}$:
\[ L_{12}=(s_1\partial_{s_2}-s_2\partial_{s_1})^2+\frac{a_3s_1^2}{s_2^2}+\frac{a_2s_2^2}{s_1^2},\]
\[ L_{13}=(s_1\partial_{s_3}-s_3\partial_{s_1})^2+\frac{a_2s_3^2}{s_1^2},\
L_{23}=(s_2\partial_{s_3}-s_3\partial_{s_2})^2+\frac{a_3s_3^2}{s_2^2},\]
\[ L=\{(s_1\partial_{s_3}-s_3\partial_{s_1}),(s_1\partial_{s_4}-s_4\partial_{s_1})\}
+\{(s_2\partial_{s_3}-s_3\partial_{s_2}),(s_2\partial_{s_4}-s_4\partial_{s_2})\} \]
\[ -\frac{a_1s_3}{\sqrt{s_1^2+s_2^2+s_3^2}}-2s_3s_4(\frac{a_2}{s_1^2}+\frac{a_3}{s_2^2}).\]
The conformally St\"ackel equivalent flat space system is
\[H_{flat}= \partial_x^2+\partial_y^2+\partial_z^2+\frac{2a_1(1-r^2)}{r(r^2+1)^2}+\frac{a_2}{x^2}+\frac{a_3}{y^2}+\frac{4a_4}{(r^2+1)^2}.\]
Here $r^2=x^2+y^2+z^2$.
This 3-parameter system extends to the  4th order 4-parameter system $ H'_{coulsphere}\Theta=0$ where 
\be \label{H'coulsphere} H'_{coulsphere}=\Delta_S+
\frac{a_0s_4}{\sqrt{s_1^2+s_2^2+s_3^2}}+\frac{a_1}{s_1^2}+\frac{a_2}{s_2^2}+\frac{a_3}{s_3^2}+a_4,\ee
and $ \Delta_S=\sum_{1\le j<k\le 4}J_{jk}^2$
is the Laplace-Beltrami operator on the 3-sphere. Here $J_{jk}=s_j\partial_{s_k}-s_k\partial_{s_j}$.
A basis of 2nd and 4th order symmetries is $\{H'_{coulsphere}, L_{12},L_{13},L_{23},K_1,K_3\}$ with
\[ L_{12}=J_{12}^2+\frac{a_2s_1^2}{s_2^2}+\frac{a_1s_2^2}{s_1^2},\
 L_{13}=J_{13}^2+\frac{a_1s_3^2}{s_1^2}+\frac{a_3s^2_1}{s^2_3},\
L_{23}=J_{23}^2+\frac{a_2s_3^2}{s_2^2}+\frac{a_3s_2^2}{s_3^2},\]
\[K_1=-4M_1^2-\frac12\left\{ \left(\{s_1,J_{41}\}+\{s_2,J_{42}\}+\{s_3,J_{43}\}\right)^2,\frac{a_3}{s_3^2}\right\}
+a_3\frac{5-7(s_1^2+s_2^2)}{s_3^2}, \]\
\[K_3=-4M_3^2-\frac12\left\{ \left(\{s_1,J_{41}\}+\{s_2,J_{42}\}+\{s_3,J_{43}\}\right)^2,\frac{a_1}{s_1^2}\right\}
+a_1\frac{5-7(s_3^2+s_2^2)}{s_1^2}, \]\
\[ M_1=\frac12\left(\{J_{23},J_{42}\}+\{J_{13},J_{41}\}\right)-\frac{a_0}{2}\frac{s_3}{\sqrt{s_1^2+s_2^2+s_3^2}}
 -a_1\frac{s_3s_4}{s_1^2}-a_2\frac{s_3s_4}{s_2^2}-a_3\frac{s_4}{s_3},\]
 \[ M_3=\frac12\left(\{J_{21},J_{42}\}+\{J_{31},J_{43}\}\right)-\frac{a_0}{2}\frac{s_1}{\sqrt{s_1^2+s_2^2+s_3^2}}
 -a_3\frac{s_1s_4}{s_3^2}-a_2\frac{s_1s_4}{s_2^2}-a_1\frac{s_4}{s_1}.\]
Note that for $a_3=0$, $K_1$ becomes a perfect square.
The conformally St\"ackel equivalent flat space system is
\[H'_{flat}= \partial_x^2+\partial_y^2+\partial_z^2+\frac{2a_0(1-r^2)}{r(r^2+1)^2}+\frac{a_1}{x^2}+\frac{a_2}{y^2}+\frac{a_3}{z^2}+\frac{4a_4}{(r^2+1)^2}.\]
Here $r^2=x^2+y^2+z^2$. We will show that this system contracts to the extended Coulomb system in flat space. Though we have no proof, there is evidence that it is not a B\^ocher
contraction of another such system. Thus it is a natural candidate for producing a family of similar systems by B\^ocher contraction from this single source. 
We expect that the 6 generators determine a closed algebra but we have not carried out the formidable calculations to verify this.

Our procedure will be to construct many more examples of 3-parameter 2nd order systems that extend to 4-parameter 4th order systems by applying
all possible B\^ocher contractions to (\ref{Hcoulsphere}) and (\ref{H'coulsphere}). 

In \S \ref{Stackel} we review the action of the St\"ackel transform on 3D Helmholtz superintegrable systems and in \S \ref{laplace} we relate Helmholtz and Laplace superintegrable systems.
In \S \ref{Bocher} we introduce B\^ocher contractions of 3D Laplace systems and determine their basic properties. In \S \ref{Nondeg} we 
review the conformally superintegrable nondegenerate Laplace systems and describe their relationship to B\^ocher contractions 
and St\"ackel transforms. The next two sections contain our basic detailed results. In \S \ref{Semideg} we list all semidegenerate 
conformally superintegrable Laplace systems that can be obtained via sequences of special B\^ocher contractions of system
(\ref{Hcoulsphere}). In \S \ref{4thordera} we list all 4th order  
conformally superintegrable Laplace systems that can be obtained via sequences of special B\^ocher contractions of system
(\ref{H'coulsphere}) and extend at least one semidegenerate system. We conclude with some remarks on unsolved problems.

\section{The St\"ackel transform on 3D manifolds}\label{Stackel}
For a conformally flat manifold with metric
$ds^2=\lambda (x,y,z)(dx^2+dy^2+dz^2)$  in Cartesian-like coordinates, a formally self-adjoint  Hamiltonian operator takes the form
\be \label{hath}
 H=\frac{1}{\lambda^{3/2}}\sum_{k,j=1}^3\partial_k\left(\delta^{kj}\lambda^{1/2}\partial_j\right)+V(x,y,z).
\ee Here  $\delta^{kj}$ is the Kronecker
delta, the measure is $\lambda^{3/2} \ dx\ dy\ dz$ and we assume all boundary terms are zero. Without loss of generality, we can assume that all even-order
symmetry operators for $H$ are formally self-adjoint and all odd order symmetries are formally skew-adjoint.

We can perform a gauge transformation ${\hat H}=e^{{\cal R}}{ H}e^{-{\cal R}}$ such that $\hat H$ is more suitable for St\"ackel transforms.
We  choose $\cal R$ such that the
differential operator part of $\hat H$ is formally self-adjoint with respect to the measure $\lambda\ dx\ dy\ dz$.
It is straightforward to show that  if we set ${\cal R}=\frac14\ln \lambda$, we have
\be {\hat H}=e^{{\cal R}}{ H}e^{-{\cal
  R}}=\frac{1}{\lambda}\sum_{i=1}^3\left(\partial_{ii}-{\cal R}_{ii}+{\cal R}_i^2\right)+V
 =\frac{1}{\lambda}\sum_{i=1}^3\left(\partial_{ii}\right)+{\hat V}.\ee
Here $V=-\frac{R}{8}+{\hat V}$ where $R$ is the scalar curvature. Thus the modified potential merely changes by an additive constant for a constant
curvature space but is nontrivial for other spaces.

The 3D quantum St\"ackel transform of a superintegrable conformally flat system $H=\Delta_3+V$ was defined in \cite{Laplace2011}. We merely note here the
simplification achieved by using the form $\hat H$. Suppose $V= V_0(x,y,z)+\alpha U(x,y,z)$ where $U\ne 0$ and $\alpha $ is a parameter. Then $U$ determines a St\"ackel transform
of $H$ to system $\tilde H$ defined by $\hat{\tilde H}= \frac{1}{U}{\hat H}$, i.e., multiplication on the left by the function $1/U$.
To obtain $\tilde H$ explicitly we perform an inverse gauge transformation. Modulo a transposition of symmetry operators, the transform preserves
the quadratic algebra structure equations, \cite{KMS2016a}.

\section{Laplace equations}\label{laplace}
Given  the eigenvalue equation $H\Theta=E\Theta$ where $H$ is a 2nd order superintegrable system (\ref{hath}) we can associate it with a unique
Laplace equation as follows:
The eigenvalue equation is equivalent to
\be\label{eigen1}{\hat H}\Theta\equiv  \left(\frac{1}{\lambda(x,y,z)}\sum_{k=1}^3\partial^2_k+{\hat V}(x,y,z)\right)\Theta=E\Theta,\ee
which in turn is  equivalent to the Laplace equation
\be\label{laplace1} ({\tilde \Delta}+{\tilde V})\Theta\equiv \left(\sum_{k=1}^3\partial^2_k+\lambda {\hat V}
-E\lambda\right)\Theta=0,\ee
where ${\tilde \Delta}$ is the flat space Laplacian and ${\tilde V}=\lambda V-E \lambda$. Now we are considering $E$ as a parameter in the
potential $\tilde V$.

We give a more general definition of Laplace systems.
\begin{definition}
Systems of  Laplace type are of the form \be\label{Laplace} H\Theta\equiv \Delta_n\Theta+V\Theta=0.\ee
 Here $\Delta_n $ is the Laplace-Beltrami operator on a real or complex conformally flat $n$-dimensional Riemannian or pseudo Riemannian manifold.
A {\it conformal symmetry} of this equation is a partial differential operator  $ S$ in the variables ${\bf x}=(x_1,\cdots,x_n)$  such
that $[ S, H]\equiv SH-HS=R_{ S} H$ for some differential operator  $R_{S}$.
A conformal symmetry maps any solution $\Psi$ of (\ref{Laplace}) to another solution. Two conformal symmetries ${ S}, { S}'$ are
identified if $S=S'+RH$ for some differential operator $R$, since they agree on the solution space of (\ref{Laplace}).
The system is {\it conformally superintegrable} if there are $2n-1$ functionally independent conformal symmetries,
${ S}_1,\cdots,{ S}_{2n-1}$ with ${ S}_1={ H}$. It is second order conformally superintegrable if each
symmetry $S_i$ can be chosen to be a  differential operator of at most second order.
\end{definition}

{\bf Facts},  \cite{KMS2016a}:
\begin{itemize}
 \item If $S$ is a ordinary symmetry of Hamiltonian $H=\frac{1}{\lambda}H'$, i.e., $[S,H]=0$,
 where $\lambda$ is a function, then $S$ is a conformal symmetry of the Laplace equation $H'\Theta=0$.
 \item If $S$ is a conformal symmetry of Hamiltonian $H=\frac{1}{\lambda}H'$
 where $\lambda$ is a function, then $S$ is a conformal symmetry of the Laplace equation $H'\Theta=0$.
 \item If $S$ is an ordinary symmetry of the Hamiltonian $H$, ${\cal R}(x,y,z)$ is a function, and ${\hat H}=e^{-{\cal R}}He^{\cal R}$ is
 a gauge transformed Hamiltonian, then ${\hat L}=e^{-{\cal R}}Le^{\cal R}$ is an ordinary symmetry of $\hat H$.
\end{itemize}

\begin{definition} Let $n=3$. We say that conformally superintegrable system (\ref{Laplace}) is {\it nondegenerate} if the potential $V$ is
5-parameter, i,e, $V(x,y,z)=\sum_{j=1}^5a_jV^{(j)}(x,y,z)$ where the $a_j$ are arbitrary parameters and the set  $\{V^{(1)},\cdots,V^{(5)}\}$ is linearly
independent over the manifold.
\end{definition}

In analogy with St\"ackel transforms of Helmholtz systems we can define conformal St\"ackel transforms of Laplace systems. Basic facts, \cite{ERM2016}:
\begin{itemize}
 \item Conformal St\"ackel Transform (CST): Assume
 \[ H\Psi=(\partial^2_{x}+\partial^2_{y}+\partial^2_z+V(x,y,z))\Psi=0;\quad V=V_0+\alpha U,\]
 \[ {\rm metric}:\ ds^2=dx^2+dy^2+dz^2,\quad {\rm measure}:\ dx\ dy\ dz,\]
 \[ CST:\quad {\tilde H}\Psi=U^{-1}(\partial^2_{x}+\partial^2_{y}+\partial^2_z)+U^{-1}V_0+\alpha)\Psi=0,\]
 \[ {\rm metric}:\ ds^2=U(dx^2+dy^2+dz^2),\quad {\rm measure}:\ U^{3/2}\ dx\ dy\ dz.\]
\item  Transformation to self-adjoint form (SA): Set $\Psi=S\Phi$, ${\hat H}=S^{-1}{\tilde H}S$, where $S=U^{1/4}$. Then the SA form is ${\hat H}\Psi=0$ where
 \[ {\hat H}=\frac{1}{U^{\frac32}}\partial_x(U^{\frac12}\partial_x)+\frac{1}{U^{\frac32}}
 \partial_y(U^{\frac12}\partial_y) +\frac{1}{U^{\frac32}}
 \partial_z(U^{\frac12}\partial_z)-\frac18{\cal R}  +\frac{V_0}{U}+\alpha,\]
  and $\cal R$ is the Riemann scalar curvature.
  \end{itemize}

\section{B\^ocher contractions}\label{Bocher} For constant curvature Helmholtz systems the underlying manifold admits the symmetry algebra
$e(2,\C)$ (flat space), or $so(3,\C)$ (the 2-sphere). Limits of these superintegrable systems to other superintegrable systems are induced  by
generalized In\"on\"u-Wigner contractions of these Lie algebras, \cite{Wigner, KM2014}. For Helmholtz systems  on manifolds with lower  or no nontrivial
symmetry algebra at all it is not clear how to classify such limits. However all these systems are equivalent to conformally superintegrable Laplace systems
on flat space, which has conformal symmetry algebra $so(5,\C)$, the Lie algebra of the conformal group, \cite{Bocher}.  In his 1894 thesis B\^ocher
developed a geometrical method for
finding and classifying the R-separable orthogonal coordinate systems for the flat space Laplace equation $\Delta_n\Psi=0$ in $n$ dimensions (no potential).
He took advantage of the  conformal
symmetry of these equations. The conformal symmetry algebra in the complex case is $so(n+2,\C)$. We will use his ideas for $n=3$ ,
but applied to the Laplace equations with potential.

The conformal symmetry algebra of the flat space Laplacian $\partial_{x}^2+\partial_{y}^2+\partial_{z}^2$ has 10 generators:{\small
\[ \partial_{x},  \partial_{y},  \partial_{z},\quad K_1=x-(x^2+y^2+z^2)\partial_x+2x(x\partial_{x}+y\partial_y+z\partial_z),\]
\[ K_2=y-(x^2+y^2+z^2)\partial_y+2y(x\partial_{x}+y\partial_y+z\partial_z),\]
\[K_3=z-(x^2+y^2+z^2)\partial_z+2z(x\partial_{x}+y\partial_y+z\partial_z),\ D=x\partial_x+y\partial_y+z\partial_z+\frac12
,\]
\[ J_{12}=x\partial_{y}-y\partial_{x}=-J_{21},\ J_{23}=y\partial_z-z\partial_y=-J_{32},\ J_{31}=z\partial_x-x\partial_z=-J_{13},\]}
nonlinear in the $K$-operators.
B\^ocher linearizes this action through the introduction of pentaspherical coordinates on flat space. These are projective coordinates
$(x_1,\cdots,x_5)$ that satisfy
\[ x_1^2+x_2^2+x_3^2+x_4^2+x_5^2=0,\ ({\rm The \ null\ cone.})\quad \sum_{k=1}^5x_k\partial_{x_k}=0.\]

They are related to Cartesian coordinates $(x,y,z)$ and to coordinates  on the 3-sphere $(s_1,s_2,s_3,s_4)$, \ $\sum_{i=1}^4s_i^2=1$,\ 
by 
\[ \frac{x_1}{X}=\frac{x_2}{Y}=\frac{x_3}{Z}=2T,\ x_4+ix_5=-2T^2,\ x_4-ix_5=2(X^2+Y^2+Z^2).\]
\[ x=\frac{X}{T}=-\frac{x_1}{x_4+ix_5},\  y=\frac{Y}{T}=-\frac{x_2}{x_4+ix_5},\ z=\frac{Z}{T}=-\frac{x_3}{x_4+ix_5},  \]
\[ x=\frac{s_1}{1+s_4},\  y=\frac{s_2}{1+s_4},\  z=\frac{s_3}{1+s_4},\]
\[ s_1=\frac{2x}{r^2+1},\  s_2=\frac{2y}{r^2+1},\ s_3=\frac{2z}{r^2+1},\ s_4=\frac{1-r^2}{r^2+1},\ r^2=x^2+y^2+z^2,\]
\[ { H}=\partial_x^2+\partial_y^2+\partial_z^2+V_F=(x_4+ix_5)^2(\sum_{k=1}^5\partial_{x_k}^2+V_B)=(1+s_4)^2(\sum_{j=1}^4\partial_{s_j}^2+V_S),\]
 \[V_F=(x_4+ix_5)^2\,V_B=(1+s_4)^2\,V_S,\]
\[ (1+s_4)=-i\frac{(x_4+ix_5)}{x_5},\ (1+s_4)^2=-\frac{(x_4+ix_5)^2}{x_5^2},\]
\[ s_1=\frac{ix_1}{x_5},\ s_2=\frac{ix_2}{x_5},\ s_3=\frac{ix_3}{x_5},\ s_4=\frac{-ix_4}{x_5}.\]
Here $\sum_{j=1}^4\partial_{s_j}^2$ is the Laplace-Beltrami operator on the 3-sphere. Thus 
\[ \left(\partial_x^2+\partial_y^2+\partial_z^2+V_F\right)\Theta=0 \Leftrightarrow  \left(\sum_{k=1}^5\partial_{x_k}^2+V_B\right)\Theta=0
  \Leftrightarrow \left(\sum_{j=1}^4\partial_{s_j}^2+V_S\right)\Theta=0.\]

\noindent {\bf Relation to flat space and 3-sphere 1st order conformal constants of the motion}
We define
\[ L_{jk}=x_j\partial_{x_k}-x_k \partial_{x_j}, \ 1\le j,k\le 5,\ j\ne k,\]
where $L_{jk}=-L_{kj}$. Note that this is a basis for $so(5,\C)$. The generators for flat space conformal symmetries are related to these via
\[ \partial_x=L_{14}+iL_{15},\ \partial_y=L_{24}+iL_{25},\ \partial_z=L_{34}+iL_{35},\  D=iL_{45},\]
\[ J_{k\ell}=L_{k\ell},,\  K_\ell=L_{\ell 4}-iL_{\ell5},\quad k,\ell=1,2,3,\ k\ne \ell.\]

The generators for $3$-sphere conformal constants of the motion are related to the $L_{jk}$ via
\[ L_{12}=J_{12},\ L_{13}=J_{13},\ L_{23}=J_{23},\
 L_{14}=J_{41}, L_{24}=J_{42},\ L_{34}=J_{43},\]
\[ L_{15}=-i\partial_{s_1},\  L_{25}=-i\partial_{s_2},\  L_{35}=-i\partial_{s_3},\  L_{45}=i\partial_{s_4}.\]

B\^ocher introduced a prescription for taking limits of quadratic forms on the null cone which lead to the construction of all orthogonal separable coordinates
for the flat space free Laplace, wave and Helmholtz equations. We now recognize that these limits are generalized In\"on\"u-Wigner 
contractions of the conformal Lie algebra to itself. We call them B\^ocher contractions. A formal treatment was given in \cite{EMS2016},
with an emphasis on
dimension $n=2$. Only minor modifications are needed for dimension 3 and higher:

\begin{definition}
Let 
\[ {\bf x}={\bf A}(\epsilon){\bf y},\ {\rm where}\ {\bf x}=(x_1,\cdots,x_{n+2}), {\bf y}=(y_1,\cdots,y_{n+2})\]
are column vectors, and  
${\bf A}=(A_{jk}(\epsilon))$, is an $(n+2)\times (n+2)$ matrix with matrix elements 
$ A_{kj}(\epsilon)=\sum_{\ell=-N}^Na^\ell_{kj}\epsilon^\ell$. Here $N$ is a nonnegative integer and the $a^\ell_{kj}$ are 
complex constants.
The matrix $\bf A$ defines a {\bf B\^ocher contraction} of the conformal algebra $so(n+2,\C)$ to itself provided 
\be \label{condition1}1):\quad 
\det ({\bf A})=\pm 1,\ {\rm  constant\ for\  all\ } \epsilon\ne 0,\ee 
\be\label{Bocher2} 2):\quad{\bf x}\cdot{\bf x}\equiv \sum_{j=1}^{n+2}x_i(\epsilon)^2={\bf y}\cdot{\bf y}+O(\epsilon).\ee
If, in addition, ${\bf A}\in O(n+2,\C)$ for all $\epsilon\ne 0$  the matrix $\bf A$ defines a {\bf special B\^ocher contraction}. \end{definition}
For a special B\^ocher 
contraction ${\bf x}\cdot{\bf x}={\bf y}\cdot{\bf y}$, with no $\epsilon$ error term.

This is a contraction in the generalized In\"on\"u-Wigner sense. Indeed, let $L_{ts}=x_t\partial_{x_s}-x_s\partial_{x_t}$, ${s\ne t}$ 
be a generator of $so(n+2,\C)$ and ${\bf \tilde A}(\epsilon)={\bf A}^{-1}$ be the matrix inverse. We have the expansion
\be\label{L2} L_{ts}=\sum_{k,\ell}(A_{tk}{\tilde A}_{\ell s}-A_{sk}{\tilde A}_{\ell t})y_k\partial_{y_\ell}
=\epsilon^{\alpha_{ts}}\left(\sum_{k\ell}F_{k\ell}\ y_k\partial_{y_\ell}+O(\epsilon)\right),\ee
where $\bf F$ is a constant nonzero matrix. Here the integer $\alpha_{ts}$ is the smallest power of $\epsilon$ occurring in the expansion of $L_{ts}$. Now consider the product
$L_{ts}( {\bf x}\cdot {\bf x})$. On one hand it is obvious that $L_{ts}({\bf x}\cdot {\bf x})\equiv 0$, but on the other hand  the expansions (\ref{Bocher2}),(\ref{L2}) yield
\[ L_{ts}( {\bf x}\cdot {\bf x})=\epsilon^{\alpha_{ts}}\left(\sum_{k\ell}F_{k\ell}\ y_k\partial_{y_\ell}\right)(\sum_{j}y_j^2)+O(\epsilon^{\alpha_{ts}}).\]
Thus, $\left(\sum_{k\ell}F_{k\ell}\ y_k\partial_{y_\ell}\right)(\sum_{j}y_j^2)\equiv 0$ 
for $\bf F$ a constant nonzero matrix. However, the only  differential operators of the form 
$\sum_{k\ell}F_{k\ell}\ y_k\partial_{y_\ell}$ that map ${\bf y}\cdot {\bf y}$ to zero are elements of $so(n+2,\C)$:
\[\sum_{k\ell}F_{k\ell}\ y_k\partial_{y_\ell}=\sum_{j>k}b_{jk}L'_{jk},\quad L'_{jk}=y_j\partial_{y_k}-y_k\partial_{y_j}.\]
Thus
\be\label{Bocherbasis1} \epsilon^{-\alpha_{ts}}L_{ts}=\sum_{j>k}b_{jk}L'_{jk}\equiv L'\ee
and this determines the contraction of $L_{ts}$ to $L'$. Similarly, if we apply this same procedure to the operator $L=\sum_{t>s}c(\epsilon)_{ts}L_{ts}$ for any
rational polynomials $c_{ts}(\epsilon)$
we will obtain an operator $L'=\sum_{j>k}b_{jk}L'_{jk}$ in the limit. Further, due to condition (\ref{condition1}), by choosing the $b_{ts}$ appropriately 
we can obtain any 
$L'\in so(4,\C)$ in the limit. In this sense the mapping $L\to L'$ is onto. Note that if $\bf A$ doesn't depend on $\epsilon$ then the contraction will be the identity mapping. Our interest is in the 
cases where $\bf A$ has nontrivial dependence on $\epsilon$.

\begin{theorem}\label{theorem1} Suppose the matrix ${\bf A}(\epsilon)$ defines a B\^ocher contraction of $so(n+2,\C)$. Let $\{L_{t_is_i},\ i=1,\cdots, 6\}$ be an ordered 
basis for $so(n+2,\C)$ such that $\alpha_{t_1s_1}\le \alpha_{t_2s_2}\le \cdots \le \alpha_{t_6s_6}$. Then there is an ordered basis 
$\{ L_j,\ j=1,\cdots,n+2\}$ such that
\begin{enumerate} \item $L_j\in {\rm span}\{L_{t_is_i},\ i=1,\cdots, j\}$
\item There are integers $\alpha_1\le \alpha_2\le\cdots\le\alpha_6$ such that \[\lim_{\epsilon \to 0}\frac{L_j}{\epsilon^{\alpha_j}}=L_j',\quad 1\le\ j\le 6,\]
and $\{L'_j,\ j=1,\cdots,n+2\}$ forms a basis for $so(n+2,\C)$ in the $y_k$ variables.
\end{enumerate}
\end{theorem}
\noindent {\bf Proof}: Induction on $j$. For $j=1$ the result follows from (\ref{Bocherbasis1}). 
Assume the assertion is true for $j\le j_0<n+2$. Then, due to the nonsingularity condition (\ref{condition1}), we can always find  polynomials in $\epsilon$, $\{a_1(\epsilon),a_2(\epsilon),\cdots, a_{j_0}(\epsilon)\}$ such that 
\[ L_{j_0+1}=L_{t_{j_0+1},s_{j_0+1}}-\sum_{i=1}^{j_0} a_iL_i=\epsilon^{\alpha_{j_0+1}}L'_{j_0+1}+O(\epsilon^{\alpha_{j_0+2}}),\]
where $L'_{j_0+1}$ is linearly independent of $\{L_i',\ 1\le i\le j_0\}$ and $\alpha_{j_0+1}\ge \alpha_{j_0}$. $\Box$
 
 Note that the proof applies to quadratic forms in general. For the definition and proof, the null cone condition ${\bf x}\cdot{\bf x}=0$ is never imposed.

\medskip

Just as in \cite{EMS2016}, we can compose two B\^ocher contractions ${\bf A}(\epsilon_1)$ and  ${\bf B}(\epsilon_2)$ and obtain another B\^ocher contraction, 
though in general not uniquely. However, if $\bf A,B$ are special B\^ocher contractions then composition is just matrix multiplication within the group $O(n+2,\C)$,
uniquely defined: we can let $\epsilon_1$ and $\epsilon_2$ go to zero independently and obtain the same contraction limit.

 \subsection{Special B\^ocher contractions}
Special B\^ocher contractions are much easier to understand and manipulate than general B\^ocher contractions: composition is merely matrix
multiplication. The contractions that arise from the B\^ocher recipe are not ``special''. However,  just as in \cite{EMS2016} for $n=2$, in the case $n=3$  we can 
associate a special B\^ocher contraction with each contraction obtained from B\^ocher's recipe, such that the special 
contraction contains the same basic geometrical information. 

Extending constructions due to Jacobi and Liouville for obtaining orthogonal separable coordinates for the free Helmholtz equations in Euclidean $n$-space and the $n$-sphere,
B\^ocher showed that (choosing $n=3$ for the purposes of this paper) that for given $x,y,z$ the pair of quadratic forms
\[ \Omega\equiv\sum_{i=1}^{5}x_i^2=0,\quad \Phi\equiv\sum_{i=1}^{5}\frac{x_i^2}{\lambda -e_i}=0,\]
where $\Omega=0$ is the null cone, determines 5 solutions $\lambda=y^1,\cdots,y^5$ and the $y^j$ are orthogonal cyclidic R-separable coordinates for the free Laplace 
equation $(\partial_x^2+\partial_y^2+\partial_z^2)\Theta=0$. Here, the $e_i$ are pairwise distinct constants. B\^ocher observed that the two quadratic forms 
$\Omega$ and $\Phi$ are such that $\Phi$ has  elementary divisors $ [1\cdots 1]$ relative to the form $\Omega$. 
(In other words, we can consider the quadratic forms as $5\times 5$ symmetric matrices, diagonal in this case. 
Here $\Omega$ corresponds to the identity matrix. The $[1,\cdots,1]$ notation refers to the fact that the $5$ eigenvalues $1/(\lambda -e_i)$ of the
$\Phi$ matrix with respect to $\Omega$ are pairwise distinct.) In fact if we have two quadratic forms related in this way they could be written more generally as
\be\label{2.23s} \Omega=\sum_{i=1}^{5}a_iz_i^2,\quad \Phi=\sum_{i=1}^{5}a_i\lambda_iz_i^2,\ee where $x_i^2=a_iz_i^2$ and the $a_i$ are nonzero constants. 
(Note that the $\lambda_i$ are the eigenvalues of $\Phi$ with respect to $\Omega$.) If exactly 2 of the eigenvalues are equal the elementary divisors are denoted 
$[2,1,1,1]$. Similarly the other possible elementary divisors are $[2,2],[311],[41]$ and $[5]$, where $[5]$ corresponds to $\lambda_1=\lambda_2=\cdots =\lambda_5$. 
B\^ocher showed that a family of orthogonal R-separable coordinates for the Laplace equation could be associated to each of these 6 elementary divisors. 
Moreover, B\^ocher provided a recipe $x_i(\epsilon),  \lambda_i(\epsilon)$, such that the coordinates and the defining quadratic forms for each
of the elementary divisors $[n_1n_2n_3]$ could be obtained in the limit as $\epsilon\to 0$. In \cite{EMS2016} it was observed that each of B\^ocher's recipes $x_i(\epsilon)$
defined a B\^ocher contraction and by specializing their adjustable parameters we could obtain the  ``special''  B\^ocher contractions. An important advance in recognizing 
B\^ocher's recipes as contractions is that they are applicable to any superintegrable system, not just to $[11111]$.

A more general way to construct special B\^ocher contractions is to make use of the normal forms for conjugacy classes of $so(5,\C)$ under the adjoint action of 
$SO(5,\C)$, as derived in  \cite{Gant}. This was discussed in \cite{EMS2016} for the case $n=2$  and the extension to $n=3$ is straight-forward. Except for the contraction $[11111]\downarrow [5]$ 
the new contractions follow easily from the $n=2$ results. For the remaining contraction, the result is a special case of B\^ocher's prescription.
The results are as follows:

\begin{enumerate}
 \item Contraction $[11111]\downarrow  [2111]$,
\[
x_1 \ = \ \frac{\left(1+\epsilon ^2\right) y_1}{2 \epsilon }-\frac{i \left(-1+\epsilon ^2\right) y_2}{2 \epsilon } \ ,  
  x_2 \ = \ \frac{i \left(-1+\epsilon ^2\right) y_1}{2 \epsilon }+\frac{\left(1+\epsilon ^2\right) y_2}{2 \epsilon }\]
 \[ x_3 \ = \  y_3    \ ,  
  x_4 \ = \  y_4    \ , 
  x_5 \ = \  y_5    \ .\]
\item Contraction $[11111]\downarrow  [221]$,
\[
 x_1 \ = \ \frac{\left(1+\epsilon ^2\right) y_1}{2 \epsilon }-\frac{i \left(-1+\epsilon ^2\right) y_2}{2 \epsilon } \ , 
  x_2 \ = \  \frac{i \left(-1+\epsilon ^2\right) y_1}{2 \epsilon }+\frac{\left(1+\epsilon ^2\right) y_2}{2 \epsilon }  \]
\[ x_3 \ = \  \frac{\left(1+\epsilon ^2\right) y_3}{2 \epsilon }-\frac{i \left(-1+\epsilon ^2\right) y_4}{2 \epsilon }   \ ,  
  x_4 \ = \  \frac{i \left(-1+\epsilon ^2\right) y_3}{2 \epsilon }+\frac{\left(1+\epsilon ^2\right) y_4}{2 \epsilon }   \]
 \[ x_5 \ = \   y_5 . \]
\item Contraction $[11111]\downarrow  [311]$,
\[ x_1 \ = \ \left(1-\frac{1}{2 \epsilon ^2}\right) y_1+\frac{y_2}{\epsilon }+\frac{i y_3}{2 \epsilon ^2} \ ,  
  x_2 \ = \  -\frac{y_1}{\epsilon }+y_2+\frac{i y_3}{\epsilon },\]
\[ x_3 \ = \  \frac{i y_1}{2 \epsilon ^2}-\frac{i y_2}{\epsilon }+\left(1+\frac{1}{2 \epsilon ^2}\right) y_3,\ 
 x_4 \ = \   y_4   \ ,
  x_5 \ = \   y_5.\]

\item Contraction $[11111]\downarrow  [32]$,
\[
x_1 \ = \ \left(1-\frac{1}{2 \epsilon ^2}\right) y_1+\frac{y_2}{\epsilon }+\frac{i y_3}{2 \epsilon ^2}  \ , 
 x_2 \ = \  -\frac{y_1}{\epsilon }+y_2+\frac{i y_3}{\epsilon }    \ , \]
\[  x_3 \ = \  \frac{i y_1}{2 \epsilon ^2}-\frac{i y_2}{\epsilon }+\left(1+\frac{1}{2 \epsilon ^2}\right) y_3    \ ,  
 x_4 \ = \  \frac{\left(1+\epsilon ^2\right) y_4}{2 \epsilon }-\frac{i \left(-1+\epsilon ^2\right) y_5}{2 \epsilon }    \ ,\]
\[ x_5 \ = \   \frac{i \left(-1+\epsilon ^2\right) y_4}{2 \epsilon }+\frac{\left(1+\epsilon ^2\right) y_5}{2 \epsilon }.\]

\item Contraction $[11111]\downarrow  [41]$,
\begin{equation*}
\begin{aligned}
&  x_1 \ = \ y_1+\left(\frac{1}{2 \epsilon ^2}+\frac{1}{2 \epsilon }\right) y_2+i\,\left(\frac{1}{2 \epsilon ^2}+\frac{1}{2 \epsilon }\right) y_3 \ , \\
&  x_2 \ = \ \left(-\frac{1}{2 \epsilon ^2}-\frac{1}{2 \epsilon }\right) y_1+y_2+i\,\left(\frac{1}{2 \epsilon ^2}+\frac{1}{2 \epsilon }\right) y_4     \ ,  \\
&  x_3 \ = \  -i\,\left(\frac{1}{2 \epsilon ^2}+\frac{1}{2 \epsilon }\right) y_1+y_3+\left(-\frac{1}{2 \epsilon ^2}-\frac{1}{2 \epsilon }\right) y_4    \ ,  \\
&  x_4 \ = \   -i\,\left(\frac{1}{2 \epsilon ^2}+\frac{1}{2 \epsilon }\right) y_2+\left(\frac{1}{2 \epsilon ^2}+\frac{1}{2 \epsilon }\right) y_3+y_4   \ ,  \\
&  x_5 \ = \    y_5  \ .
\label{}
\end{aligned}
\end{equation*}
\item Contraction $[11111]\downarrow  [5]$,
\begin{equation*}
\begin{aligned}
&  x_1 \ = \   \frac{y_1+\epsilon\,y_2+\epsilon^2\,y_3+\epsilon^3\,y_4+\epsilon^4\,y_5}{\sqrt{5}\,\epsilon^2}   \ , \\  &
x_2 \ = \  \frac{y_1+Z\,\epsilon\,y_2+Z^2\,\epsilon^2\,y_3+Z^3\,\epsilon^3\,y_4+Z^4\epsilon^4\,y_5}{\sqrt{5}\,Z^2\,\epsilon^2}   \ , \\ &
x_3 \ = \  \frac{y_1+Z^2\,\epsilon\,y_2+Z^4\,\epsilon^2\,y_3+Z^6\,\epsilon^3\,y_4+Z^8\epsilon^4\,y_5}{\sqrt{5}\,Z^4\,\epsilon^2} \ , \\ &
x_4 \ = \  \frac{y_1+Z^3\,\epsilon\,y_2+Z^6\,\epsilon^2\,y_3+Z^9\,\epsilon^3\,y_4+Z^{12}\epsilon^4\,y_5}{\sqrt{5}\,Z\,\epsilon^2} \ , \\ &
x_5 \ = \  \frac{y_1+Z^4\,\epsilon\,y_2+Z^8\,\epsilon^2\,y_3+Z^{12}\,\epsilon^3\,y_4+Z^{16}\epsilon^4\,y_5}{\sqrt{5}\,Z^3\,\epsilon^2} \ , \\ &
\ .
\label{}
\end{aligned}
\end{equation*}
where $Z$ is a primitive fifth root of unity: $1+Z+Z^2+Z^3+Z^4=0$.
\end{enumerate}
\subsection{Application of the B\^ocher contraction}\label{Bochersub}
Suppose we have a conformal superintegrable system of some order
\be \label{Laplacesemi} (\sum_{i=1}^{n+2}\partial_{x_i}^2+V_B)\Theta=0,\quad V_B({\bf x})=\sum_{j=1}^ka_jV^{(j)}({\bf x}),\ee
where the $a_j$ are the independent parameters in the potential and the set $\{V^{(1)},\cdots, V^{(k)}\}$ is functionally independent and parameter free.
Let  ${\bf a}=(a_1,\cdots ,a_k)$ and let 
${\bf x}={\bf A}(\epsilon){\bf y}$ be a special B\^ocher contraction of  $so(n+2,\C)$. We will show that the application of this contraction to the Laplace equation 
(\ref{Laplacesemi}) yields a unique finite limit once we determine rational functions $a_k(\epsilon)$ appropriately. Since ${\bf A}(\epsilon){\bf y}\in O(n+2)$ 
for all $\epsilon\ne 0$ it is clear that $\sum_{i=1}^{n+2}\partial_{x_i}^2\to \sum_{i=1}^{n+2}\partial_{y_i}^2$ as $\epsilon\to 0$, so we only need to show
that $V_B({\bf x}(\epsilon))\to {\hat V}_B({\bf y})$ as $\epsilon\to 0$ for appropriate $a_j(\epsilon)$. We can expand the potential as a Laurent series in $\epsilon$:
\be\label{epsilonexpansion} V_B({\bf x}(\epsilon))=\epsilon^{\alpha_1}\sum_s {\tilde a}^{(1)}_sf_s^{(1)}({\bf y})
+\epsilon^{\alpha_2}\sum_s {\tilde a}^{(2)}_sf_s^{(2)}({\bf y})+\cdots\ee
Here, $\alpha_1<\alpha_2<\alpha_3\cdots$, the parameters ${\tilde a}^{(j)}_s$ are linear combinations of the parameters $a_1,\cdots,a_k$, for each fixed $j$ \
the set $\{ f_s^{(j)}({\bf y})\}$ is functionally independent, and ${\tilde a}^{(j)}_s={\bf a}\cdot{\bf c}^{(j)}_s$ where ${\bf c}^{(j)}_s$is a nonzero $k$-vector of constants.  (At this point we impose the null cone condition  ${\bf y}\cdot{\bf y}=0$.) We order the vectors as
\[ {\bf c}^{(1)}_1,{\bf c}^{(1)}_2,\cdots,  {\bf c}^{(2)}_1,{\bf c}^{(2)}_2,\cdots \]
Let $\hat k$ be the dimension of the space spanned by these vectors. Starting with ${\bf c}^{(1)}_1= {\bf c}^{(\beta_1)}_{\gamma_1}$, choose vectors $ {\bf c}^{(\beta_\ell)}_{\gamma_\ell}$ in increasing order such that each of 
the sets $\{ {\bf c}^{(\beta_1)}_{\gamma_1}, {\bf c}^{(\beta_2)}_{\gamma_2},\cdots
{\bf c}^{(\beta_m)}_{\gamma_m}\}$ is linearly independent for $m=1,\cdots, d$.
To obtain a finite limit, we require ${\bf a}\cdot  {\bf c}^{(\beta_\ell)}_{\gamma_\ell}=b_\ell \epsilon^{-\alpha_{\beta_\ell}}$ for each $\ell=1,\cdots,d$, where the $b_\ell$ are $\epsilon$-independent parameters. It follows that 
\be\label{Epsilonexpansion}\sum_{i=1}^{n+2}\partial_{x_i}^2+ V_B({\bf x}(\epsilon))=\sum_{i=1}^{n+2}\partial_{y_i}^2+V_B({\bf y},{\bf b})+O(\epsilon),\ee
where ${\bf b}=(b_a,\cdots,b_d)$.

Now we examine the behavior of the symmetry operators under special B\^ocher contraction. The analysis is very similar to that used to prove Theorem \ref{theorem1}.
Let $\{ S_1, S_2,\cdots S_h\}$  be an ordered 
basis for the symmetries of system (\ref{Laplacesemi}).  Then there is an ordered basis 
of symmetries  $\{  S'_1(\epsilon), {S'}_2(\epsilon),\cdots { S'}_h(\epsilon)\}$ of (\ref{Laplacesemi}) for each $\epsilon\ne 0$
 such that 
\begin{enumerate} \item ${ S'}_j\in {\rm span}\{S_{i},\ i=1,\cdots, j\}$
\item There are integers $\beta_1\le \beta_2\le\cdots\le\beta_h$ such that
 \[\lim_{\epsilon \to 0}\frac{S'_j}{\epsilon^{\beta_j}}={\hat S}_j,\quad 1\le\ j\le h,\]
where $\{ {\hat S}_1,\cdots ,{\hat S}_h\}$ is a linearly independent set of operators for the contracted system
$\left(\sum_{i=1}^{n+2}\partial_{y_i}^2+{\hat V}_B({\bf y},{\bf b})\right)\Theta=0$
\end{enumerate}
Indeed applying a B\^ocher contraction ${\bf x}={\bf A}(\epsilon){\bf y}$ to a nonzero symmetry $S$ of   (\ref{Laplacesemi}) we have 
\[ S({\bf x}(\epsilon))=\epsilon^\beta\left({\hat  S}({\bf y})+O(\epsilon)\right),\]
where $\beta$ is the smallest power of $\epsilon$ occurring in the expansion of $S$. Thus 
$ \epsilon^{-\beta}S\to {\hat S}\ne 0$
as $\epsilon\to 0$. The rest is by induction on $j$. For $j=1$ the result follows. 
Assume the assertion is true for $j\le j_0<h$. Then there  are polynomials in $\epsilon$, $\{a_1(\epsilon),a_2(\epsilon),\cdots, a_{j_0}(\epsilon)\}$ such that 
\[ S'_{j_0+1}=S_{t_{j_0+1}}-\sum_{i=1}^{j_0} a_iS_i=\epsilon^{\beta_{j_0+1}}{\hat S}_{j_0+1}+O(\epsilon^{\beta_{j_0+2}}),\]
 ${\hat S}_{j_0+1}$ is linearly independent of $\{{\hat S}_i,\ 1\le i\le j_0\}$ and $\beta_{j_0+1}\ge \beta_{j_0}$.

It remains to show that the $\{{\hat S}_j\}$ are symmetries of system 
$(\sum_{i=1}^{n+2}\partial_{y_i}^2+V_B({\bf y},{\bf b})\Theta=0$.
For this, note that 
 \[\left [\sum_{i=1}^{n+2}\partial_{x_i}^2+V_B({\bf x}(\epsilon)), \epsilon^{-\beta_j}S'_j\right]
=\left[\sum_{i=1}^{n+2}\partial_{y_i}^2+{\hat V}_B({\bf y},{\bf b}), {\hat S}_j\right]+O(\epsilon).\]
The quantity on the left is $0$ for all $\epsilon\ne 0$, so the quantity on the right must vanish in the limit as $\epsilon\to 0$.
\section{3D Nondegenerate Laplace equations} \label{Nondeg}
\[ (\partial^2_{x}+\partial^2_{y}+\partial^2_z+V(x,y,z))\Psi=
(x_4+ix_5)^2\left(\sum_{k=1}^5\partial_{x_k}^2+{\tilde V}(x_1,\cdots,x_5)\right)=0,\]
where ${\tilde V}={V}/{(x_4+ix_5)^2}$.
There are 10 equivalence classes \cite{CKP2015}:
\begin{enumerate}
 \item $V_{[1,1,1,1,1]}= \frac{a_1}{x^2}+\frac{a_2}{y^2}+\frac{a_3}{z^2}+\frac{4a_4}{(x^2+y^2+z^2-1)^2}-\frac{4a_5}{(x^2+y^2+z^2+1)^2}$,
  \item $ V_{[2,1,1,1,]}=\frac{a_1}{x^2}+\frac{a_2}{y^2}+\frac{a_3}{z^2}+a_4(x^2+y^2+z^2)+a_5$,
\item $ V_{[2,2,1]}=\frac{a_1}{z^2}+\frac{a_2}{(x+iy)^2}+\frac{a_3(x-iy)}{(x+iy)^3}+a_4(x^2+y^2+z^2)+a_5$.
\item $V_{[3,1,1]}=a_1x+\frac{a_2}{y^2}+\frac{a_3}{z^2}+a_4(4x^2+y^2+z^2)+a_5$.
\item $V_{[3,2]}=a_1x+\frac{a_2}{(y+iz)^2}+\frac{a_3(y-iz)}{(y+iz)^3}+a_4(4x^2+y^2+z^2)+a_5$,
 St\"ackel equivalent to $ V_{[2,3]}=\frac{a_1}{(x+iy)^2}+\frac{a_2z}{(x+iy)^3}+a_3(x^2+y^2+z^2)+\frac{a_4(x^2+y^2-3z^2)}{(x+iy)^4}+a_5$,
\item $V_{[4,1]}=\frac{a_1}{z^2}+a_2(x-iy)+a_3\left(2(x+iy)-3(x-iy)^2\right)+\break a_4\left(z^2-2(x-iy)^3+4(x^2+y^2)\right)+a_5$,
\item $ V_{[5]}=-a_1(x+iy)+a_2(3(x+iy)^2+z)+a_3\left(16(x+iy)^3+(x-iy)+\right.$
$\left.12(x+iy)z\right) +a_4\left(5(x+iy)^4+(x^2+y^2+z^2)+6(x+iy)^2z\right)+a_5$,
\item $ V_{[0]}=a_1x+a_2y+a_3z+a_4(x^2+y^2+z^2)+a_5$,
\item $V_{[0,0]}=a_1x+a_2y+\frac{a_3}{z^2}+a_4(4x^2+4y^2+z^2)+a_5$,
\item $ V_{[A]}=a_1z+a_2(x-iy)+a_3\left((x-iy)^2+2(x+iy)\right)+a_4\left((x-iy)^3\right.$ $\left.+6(x^2+y^2+z^2)\right)+a_5$,
\end{enumerate}
It is an easy extension of 2D theory, \cite{KM2014,KMS2016a},  to show that each  B\^ocher contraction can be applied to any of these 3D nondegenerate Laplace systems 
and a superintegrable system results. (However, a functionally linearly independent system may contract to a functionally linearly dependent system.)
It has been shown that each Laplace equation can be obtained as a B\^ocher contraction of system $V[11111]$, \cite{CKP2015}, but the full contraction scheme has not yet 
been worked out.

\medskip

\section{Semidegenerate Laplace systems}\label{Semideg}
\begin{figure}[p]
\centerline{\includegraphics[width=15cm]{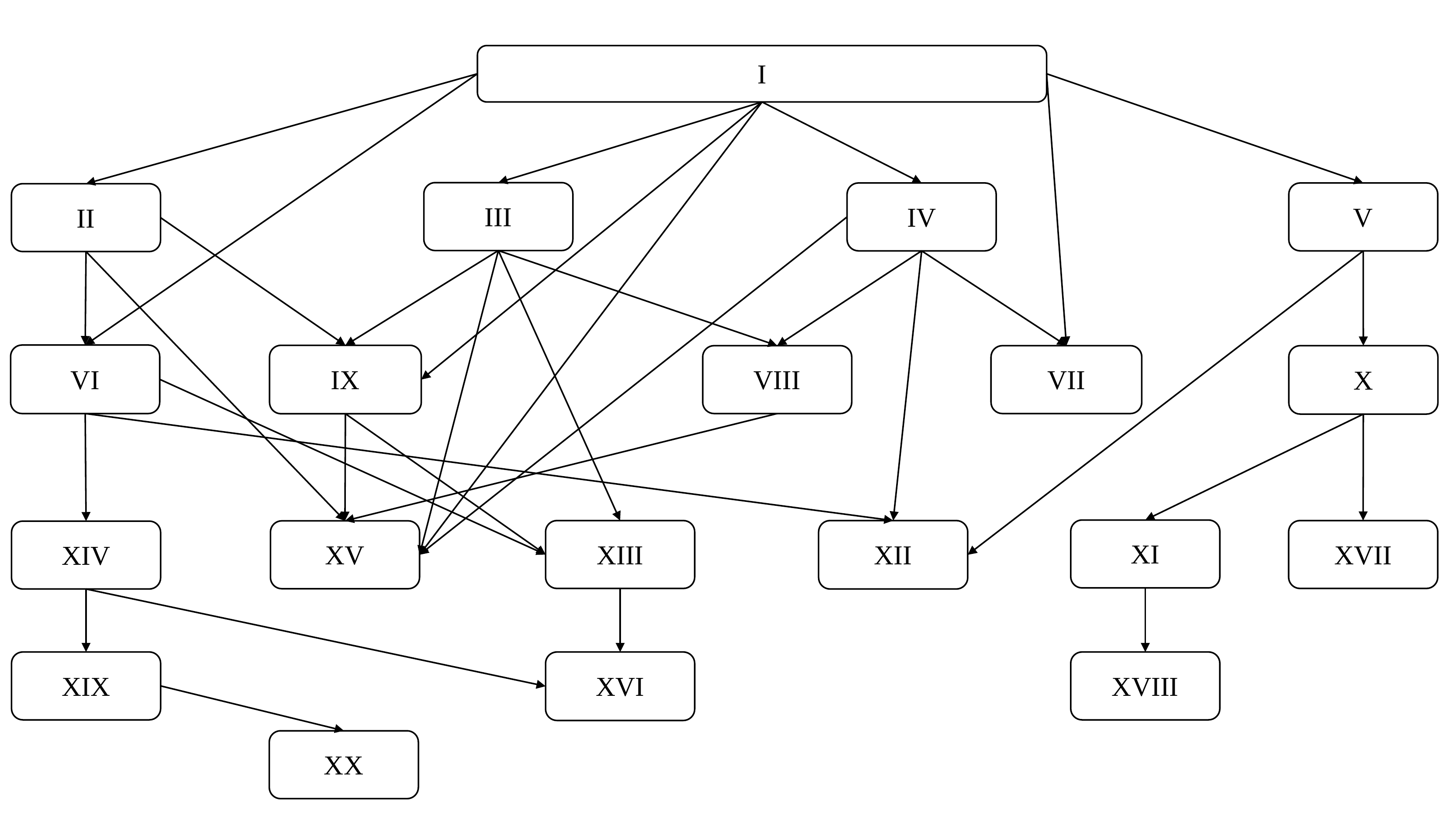}}
\caption{Contractions of semidegenerate systems. System $B$ is a B\^ocher contraction of system $A$ provided there 
is an arrow pointed from $A$ to $B$.} \label{2ndorder}
\end{figure}

Here, we designate the Laplace systems in B\^ocher and flat space Cartesian coordinates by
\[ { H}=\partial_x^2+\partial_y^2+\partial_z^2+V_F=(x_4+ix_5)^2\,(\sum_{k=1}^5\partial_{x_k}^2+V_B)=(1+s_4)^2(\sum_{j=1}^4\partial_{s_j}^2+V_S),\]
where \[V_F=(x_4+ix_5)^2\,V_B=(1+s_4)^2\,V_S.\]
We start with the ``generic'' semidegenerate system (\ref{Hcoulsphere}) and apply each B\^ocher contraction to this system as described in 
subsection \ref{Bochersub}. In this case we have $n=3,h=5,k=4$. There are 5 basic B\^ocher contractions, but each contraction is not symmetric in the coordinates $x_i$
so there are potentially $5!\times 5=120$ limits to take, though in practice this can be reduced substantially. Each contraction yields a superintegrable system,
but it need not be semidegenerate. The contracted system will have 5 independent symmetries and $d$-parameter potential. If $d<4$ then the contraction cannot cover
a full semidegenerate system so we do not count it. If $d=k=4$ but the contracted potential is functionally dependent, again the contraction cannot be 
semidegenerate. If $d=4$ and the contracted potential is functionally independent and the contracted system admits 6 linearly 
independent symmetries, then it can be extended to 
a 2nd order system with nondegenerate 4-parameter potential and is not semidegenerate. The remaining cases are semidegenerate. However we ignore
`identity``  contractions of  (\ref{Hcoulsphere}) to itself. Once we have determined
all new semidegenerate systems resulting from B\^ocher contractions of (\ref{Hcoulsphere}) we repeat the procedure for each of these
new systems. We continue this process on the results until no new semidegenerate systems appear. The process is relatively straightforward
but lengthy. Here and for the 4th order extensions we  write  the parameters in order, i.e., $V_F =\sum_j a_jV^{(j)}$, where $V_F$ is the potential in flat space and Cartesian coordinates.
We list the results:
\begin{enumerate}
 \item System I:\quad  $ V_F=$
\[-\frac{a_1(x^2+y^2+z^2-1)}{(x^2+y^2+z^2+1)^2\sqrt{x^2+y^2+z^2}}+\frac{a_2}{x^2}+\frac{a_3}{y^2}+\frac{4a_4}{(x^2+y^2+z^2+1)^2}.\]
 There are 2 constant curvature Helmholtz systems in this St\"ackel equivalence class:
 \[ \Delta_S+\frac{a_1s_4}{\sqrt{s_1^2+s_2^2+s_3^2}}+\frac{a_2}{s_1^2}+\frac{a_3}{s_2^2}+a_4,\
  \Delta_S+\frac{ia_1s_1}{s_4^2\sqrt{s_1^2+s_4^2}}-\frac{a_2}{s_2^2}+a_3+\frac{a_4}{s_4^2}.\]
 \item System II:\quad $ V_F=\frac{a_1 }{ \sqrt{x^2+y^2+ z^2}} +
\frac{a_2}{ x^2}+
\frac{a_3}{ y^2}+a_4$.\hfill\break
This is a $[11111]\downarrow  [2111]$ contraction of System I.
 There are 2 constant curvature Helmholtz systems in this St\"ackel equivalence class. One is 
 $ \Delta_F+V_F$,
 listed above, and the other is
 \[ \Delta_S-\frac{a_1}{s_2^2}+a_2+\frac{a_3}{(s_4-is_1)\sqrt{s_1^2+s_4^2}}+\frac{a_4}{(s_4-is_1)^2}.\]
 \item System III: $ V_F=a_3(x^2+y^2+z^2)+a_2-\frac{a_4}{y^2}+\frac{a_1x}{y^2\sqrt{x^2+y^2}}$.\hfill\break
This is a $[11111]\downarrow  [2111]$ and a $\downarrow  [311]$ contraction of System I.
\item System IV: $ V_F=\frac{a_3}{y^2}+a_2-\frac{a_4}{z^2}+\frac{ia_1x}{z^2\,\sqrt{x^2+z^2}}4$.\hfill\break This is a $\downarrow  [2111]$ contraction of System I.
\item System V: $ V_F=\frac{ia_1(y+iz)}{\sqrt{(y+iz)^2+1}}+a_2+\frac{a_3}{x^2}+a_4(x^2+y^2+z^2)$.\hfill\break
This is a $\downarrow  [221]$ contraction of System I.
\item System VI: $ V_F=\frac{a_3}{y^2}+a_2-a_1x-a_4(x^2+y^2+z^2)$.\hfill\break This is a $\downarrow  [2111]$ of System I and a $\downarrow  [311]$  contraction of Systems I and II.
\item System VII: $ V_F=\frac{a_1}{(x+iy)^2}+a_2(x^2+y^2+z^2)+a_3z+a_4$.\hfill\break
This is a $\downarrow  [32]$ contraction of System I and  a $\downarrow  [2111]$ contraction of System IV.
It is St\"ackel equivalent to
\[ V_F=a_1+\frac{a_2(y-iz)}{(y+iz)^3}+\frac{a_3x}{(y+iz)^3}+\frac{a_4}{(y+iz)^2}.\]
\item System VIII:  $ V_F=a_1z+a_2+\frac{a_3}{y^2}+\frac{a_4x}{y^2\sqrt{x^2+y^2}}$.\hfill\break  This is a $\downarrow  [2111]$ contraction of System III and IV, and a  $\downarrow
[311]$ contraction of Systems III and IV.
\item System IX: $ V_F=\frac{a_1}{(iy+z)\sqrt{y^2+z^2}}+a_2-a_3(x^2+y^2+z^2)+\frac{a_4}{(iy+z)^2}$.\hfill\break
 This is a $\downarrow  [221]$ contraction of System I and a $\downarrow  [2111]$ contraction of Systems II and III. It is St\"ackel equivalent to
 \[ V_F=\frac{a_1(x-iy)}{(x+iy)^3}+\frac{a_2}{(x+iy)^2}+a_3+\frac{a_4}{\sqrt{x^2+y^2+z^2}}.\]
\item System X: $V_F=a_1+\frac{a_2}{z^2} +a_3y+\frac{a_4}{\sqrt{x-iy}}$. \hfill\break
This is a $\downarrow  [2111]$ contraction of System V.
\item {System XI: $ V_F=a_1+a_2(x+iy) +\frac{a_3}{z^2}+\frac{a_4}{\sqrt{x-iy}}$.\hfill\break
 This is a $\downarrow  [2111]$ and a $\downarrow  [221]$ contraction of System X.}
\item System XII: $ V_F=a_1y+a_2+\frac{a_3}{x^2}+a_4z$.\hfill\break This is a  $\downarrow  [2111]$, and  a  $\downarrow  [311]$ contraction of Systems VI, IV and V.
\item System XIII: $ V_F=a_1+a_2x+a_3z+(x^2+y^2+4z^2)$.\hfill\break This is a  $\downarrow  [2111]$ contraction of Systems III and VI, and a  $\downarrow  [311]$ contraction of Systems VI and IX.
\item System XIV: $ V_F=a_3+a_4(x^2+y^2+z^2)+\frac{a_2}{x^2}+a_1(y+iz)$.\hfill
This is a $\downarrow  [2111]$ contraction of System VI.
\item System XV: $ V_F=\frac{a_3y}{(x+iz)^3}+\frac{a_2}{(x+iz)^2}+a_4+\frac{a_1}{\sqrt{x^2+y^2+z^2}}$.\hfill\break
This is a $\downarrow  [221]$ contraction of Systems III, IV, and VIII, a $\downarrow  [311]$  contraction of systems II and IX,
and a  $\downarrow  [32]$ contraction of System I. It is St\"ackel equivalent to
\[ V_F=a_1x +a_2+\frac{a_3}{(y+iz)^2}+\frac{a_4}{(y+iz)\sqrt{y^2+z^2}}.\]
\item System XVI: $V_F=a_1+a_2x+a_3(y+iz)+(4x^2+y^2+z^2)$. \hfill\break This is  a  $\downarrow  [311]$ contraction of System XIV,
a  $\downarrow  [2111]$ contraction of Systems XIII, and a $\downarrow  [221]$ contraction of System XIII.
\item  System XVII: $ V_F=a_1+a_2y +a_3z+\frac{a_4}{\sqrt{x-iy}}$.\hfill\break
 This is a $\downarrow  [2111]$ and a $\downarrow  [221]$   contraction of System X.
 \item  System XVIII: $ V_F=a_1+a_2(x+iy) +{a_3}{z}+\frac{a_4}{\sqrt{x-iy}}$.\hfill\break
This is a $\downarrow  [32]$ and a   $\downarrow  [41]$   contraction of System XVII,
 and  $\downarrow  [2111]$, $[221]$,$[311]$,$[32]$ and $ [41]$   contractions of System XI.  
 \item System XIX: $V_F=a_1+a_2x+a_3z+a_4\left(3(z-ix)^2+4iy\right)$.\hfill\break
This is  a  $\downarrow  [311]$ contraction of System XIV.
\item System XX: $ V_F=a_1+a_2(x-iy)+a_3(y+iz)+a_4\left(-2iz+3(y+iz)^2\right)$. This is a $\downarrow  [32]$ contraction of System XIX.
 \end{enumerate} 
Figure \ref{2ndorder} is a graphical depiction of the contraction results.
\begin{example}
 \item A functionally linearly dependent system.   This is a  $[11111]\downarrow  [311]$ contraction of Systems VI
a $[11111]\downarrow  [41]$ contraction of System I, a   $[11111]\downarrow  [2111]$ contraction of System XIX,
 and a $[11111]\downarrow  [41]$ contraction of Systems XIX,  and XX.
\[ V_F=a_1+a_2x+a_3z+a_4(x+iz)^2.\]
Note that the potential doesn't depend on $y$, so the system cannot be semidegenerate.
\end{example}
\section{Extensions of semi-degenerate Laplace systems to 4th order superintegrable systems}\label{4thordera}
To compile this list we start with the ``generic'' 4th order  system (\ref{H'coulsphere}) and apply each B\^ocher contraction to this system as described in 
subsection \ref{Bochersub}. In this case we have $n=3,h=6,k=5$. Since each of the 5 basic B\^ocher contractions is not completely 
symmetric in the coordinates $x_i$
there are potentially $5!\times 5=120$ limits to take, though again this can be reduced substantially. 
Each contraction yields a superintegrable system 
but it need not be 4th order or an extension of a semidegenerate system. The contracted system will have 6 independent symmetries
and $d$-parameter potential. If $d<5$ then the contraction cannot cover
a full 4th order  system so we do not count it. If $d=k=5$ but the contracted potential is functionally dependent, again the contraction
cannot counted. If $d=5$ and the contracted potential is functionally independent but there are  5 linearly 
independent 2nd order symmetries the contracted system  is 2nd order nondegenerate and cannot be counted. 
The remaining cases are 4th order systems with 4 2nd order symmetries. For each of these cases we must check if the system can be 
restricted to a 3-parameter system with 5 linearly independent symmetries. If so, we count it as an extension,
though we ignore
``identity``  contractions of  (\ref{H'coulsphere}) to itself. Once we have determined
all new 4th order extensions  resulting from B\^ocher contractions of (\ref{H'coulsphere}) we repeat the procedure for each of these
new systems. We continue this process on the results until no new extension systems appear. We list the results:
{\small
\begin{enumerate}
\item System i:  This is the extension (\ref{H'coulsphere}) of semi-degenerate system I.
\item System ii:   This is an extension of semi-degenerate system II.
\[ V_F=\frac{a_0 }{ \sqrt{x^2+y^2+ z^2}} +
\frac{a_1}{ x^2}+
\frac{a_2}{ y^2}+
\frac{a_3}{ z^2}+a_4.\] It  is a $\downarrow  [2111]$ contraction of System i.
\item System iii: This is an extension of Systems IV and III. It is a  $\downarrow  [2111]$ contraction of System i.
\[ V_F:=\frac{a_3}{x^2}+\frac{a_4}{y^2}+\frac{a_5z}{y^2\ \sqrt{y^2+z^2}}+a_1-a_2(x^2+y^2+z^2).\]
\item System iv: This is an extension of System V,  and a  $\downarrow  [221]$ contraction of System i.
\[ V_F=\frac{a_2}{(x+iy)^2}+a_4(x^2+y^2+z^2)+a_3+\frac{a_5(x+iy)}{\sqrt{1-(x+iy)^2}}+\frac{a_1}{z^2}.\]
\item System v: This is an extension of System VI, a  $\downarrow  [2111]$ contraction of Systems i and ii, and a
$\downarrow  [311]$ contraction of System i.
\[ V_F=\frac{a_3}{x^2}+\frac{a_1}{y^2}+a_2+a_4z-a_5(x^2+y^2+z^2).\]
\item System vi: This is an extension of System VII,  and a  $\downarrow  [221]$ contraction of Systems iii and iv.
\[ V_F=\frac{a_2}{(x+iy)^2}+a_4(x^2+y^2+z^2)+a_3+a_1z+\frac{a_5(x+iy)}{\sqrt{1-(x+iy)^2}}.\]
\item System vii: This is an extension of System VIII, a  $\downarrow  [211]$ contraction of System iii, a  $\downarrow  [221]$ contraction of System iii, and a
$\downarrow  [311]$ contraction of System  i.
\[ V_F=a_3+a_1x-a_2(4x^2+y^2+z^2)+\frac{a_4}{z^2}+\frac{a_5y}{z^2\,\sqrt{y^2+z^2}}.\]
\item System viii: This is an extension of System IX, a  $\downarrow  [2111]$ contraction of Systems ii, iii and  xvi,
and a $\downarrow  [221]$ contraction of Systems i and iii.
\[ V_F=\frac{a_3}{x^2}+\frac{a_2(y+iz)}{(y-iz)^3}+\frac{a_1}{(iy+z)^2}+\frac{a_5}{\sqrt{x^2+y^2+z^2}}+a_4.\]
It is St\"ackel equivalent to system
\[ V'_F=\frac{a_3}{z^2}-a_2(x^2+y^2+z^2)-a_1+\frac{a_5}{(x+iy)\ \sqrt{x^2+y^2}}+\frac{a_4}{(x+iy)^2}.\]
\item System ix: This is a $\downarrow  [2111]$  contraction of System iv. It is  an extension of System X.
\[ V_F=a_1+\frac{a_2}{z^2} +a_3x+a_4y+\frac{a_5}{\sqrt{x-iy}}.\]
\item System x: This is an extension of System VII, a  $\downarrow  [2111]$ contraction of Systems ii, iii and  v,
 a $\downarrow  [221]$ contraction of Systems i, ii, iii and v, a $\downarrow  [311]$ contraction of System iii,
 and  a $\downarrow  [32]$ contraction of Systems i, iii and v.
\[ V_F=\frac{a_2(x-iz)}{(x+iz)^3}+\frac{a_1}{(x+iz)^2}+a_3+a_5y-a_4(x^2+y^2+x^2).\]
It is St\"ackel equivalent to system
\[ V'_F=-a_2(x^2+y^2+z^2)+a_1+\frac{a_3}{(x+iy)^2}+\frac{a_5z}{(x+iy)^3}+\frac{a_4(x-iy)}{(x+iy)^3}.\]
\item System xi: This is an extension of Systems XI and  XII, a  $\downarrow  [2111]$ contraction of Systems iii and v,
and a $\downarrow  [311]$ contraction of Systems i, iii and v.
\[ V_F=\frac{a_3}{x^2}+a_1+a_2y-a_5(x^2+4y^2+z^2)+a_4z.\]
\item System xii: This is an extension of System XIII, a $\downarrow  [2111]$ contraction of System iv,
and  $\downarrow  [2111]$, $[221]$ and $[41]$ contractions of Systems v.
\[ V_F=\frac{a_4}{(iy+z)^3}+\frac{a_3}{x^2}+\frac{a_5(y+iz)}{(y-iz)^3}+\frac{a_2}{(iy+z)^2}+a_1.\]
St\"ackel equivalent to
\[ V'_F=a_4(x+iy)+\frac{a_3}{z^2}+a_5(x^2+y^2+z^2)+a_2+\frac{a_1}{(x+iy)^2}.\]
\item System xiii: This is an extension of System XI,  a $\downarrow  [311]$ and $[32]$ contraction of System ii.  A  $\downarrow  [221]$ contraction of Systems iii and iv,  a
  $\downarrow  [32]$ contraction of Systems i and iii, and  a $\downarrow  [2111]$ and $[32]$ contraction of System vii.
\[ V_F=a_4+\frac{a_1}{(x+iy)^2}+\frac{a_3z}{(x+iy)^3}+\frac{a_2(x^2+y^2-3z^2)}{(x+iy)^4}+\frac{a_5}{\sqrt{x^2+y^2+z^2}}.\]
It is St\"ackel equivalent to
\[ V_F=\frac{a_1}{(x+iy)^2}+a_2+a_3z+a_4(x^2+y^2+4z^2)+\frac{a_5}{(x+iy)\sqrt{x^2+y^2}}.\]
\item  System xiv: This is a $\downarrow  [41]$  contraction of Systems v and xii, and  a $\downarrow  [32]$
contraction of System xi. It is an extension of System XV.\qquad  $V_F=a_1+a_2z+$ \hfill\break 
\[ a_3(x+iy)+\frac{a_4}{(x+iy)^2}+a_5\left(3ix(z-y)+2z^2+2y^2-x^2-3xy\right).\]
\item System xv: This is a $\downarrow  [2111]$  contraction of System xiii, and
 $\downarrow  [2111]$ and $[32]$ contractions of System ix. It is an extension of System XVI.
\[ V_F=a_1+a_2x+a_3y+a_4z+\frac{a_5}{\sqrt{x-iy}}.\]
\item  System xvi: This is an extension of system XV,  a  $\downarrow  [221]$ and $[32]$ contraction of System v,
$\downarrow  [2111]$, $[221]$ and $[32]$ contractions of Systems xii, a  $\downarrow  [32]$ contraction of Systems i and iii,
$\downarrow  [2111]$, $[221]$ and $[32]$ contractions of System xii,
  $\downarrow  [2111]$  and  $[221]$ contractions of System xiv,
 and  $\downarrow  [2111]$, $[221]$  $[32]$ and $[41]$ contractions of System xi.
\[ V_F=a_1+a_2x+a_3(y+iz)+\frac{a_4}{(y+iz)^2}+a_5(4x^2+y^2+z^2).\]
\item System xvii: This is a $\downarrow  [2111]$,  $ [311]$ and $[41]$ contraction of System xvi,
a $\downarrow  [2111]$, $[221]$  $ [311]$, $[32]$ and $[41]$ contraction of Systems xiv and xv, a $\downarrow  [311]$ and $[32]$  contraction of Systems
x, v, and xxvi, a  $\downarrow  [311]$  contraction of Systems xii, vi and ix,  a  $\downarrow  [41]$  contraction of System  ix, a  $\downarrow  [32]$  contraction of System  xiii, and $\downarrow  [221]$,  $ [311]$, $[32]$ and $[41]$ contractions of System xi.
\[ V_F=a_1+a_2x+a_3y+a_4z+a_5(x+iz)^2.\] 
\end{enumerate}}

\begin{figure}[p]
\centerline{\includegraphics[width=15cm]{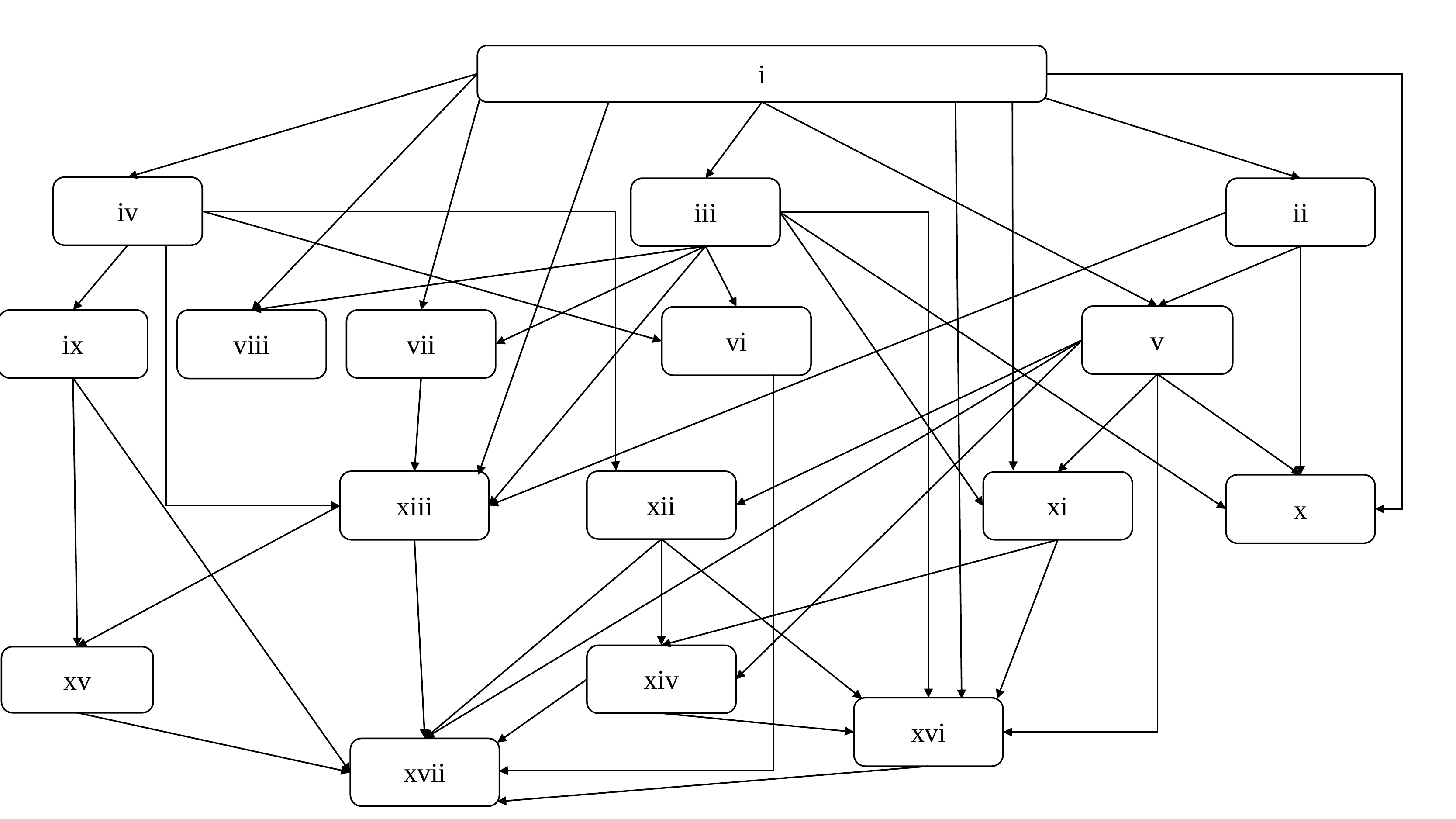}}
\caption{Contractions of 4th order systems that are extensions of semidegenerate systems} \label{4thorder}
\end{figure}

\begin{figure}[p]
\centerline{\includegraphics[width=15cm]{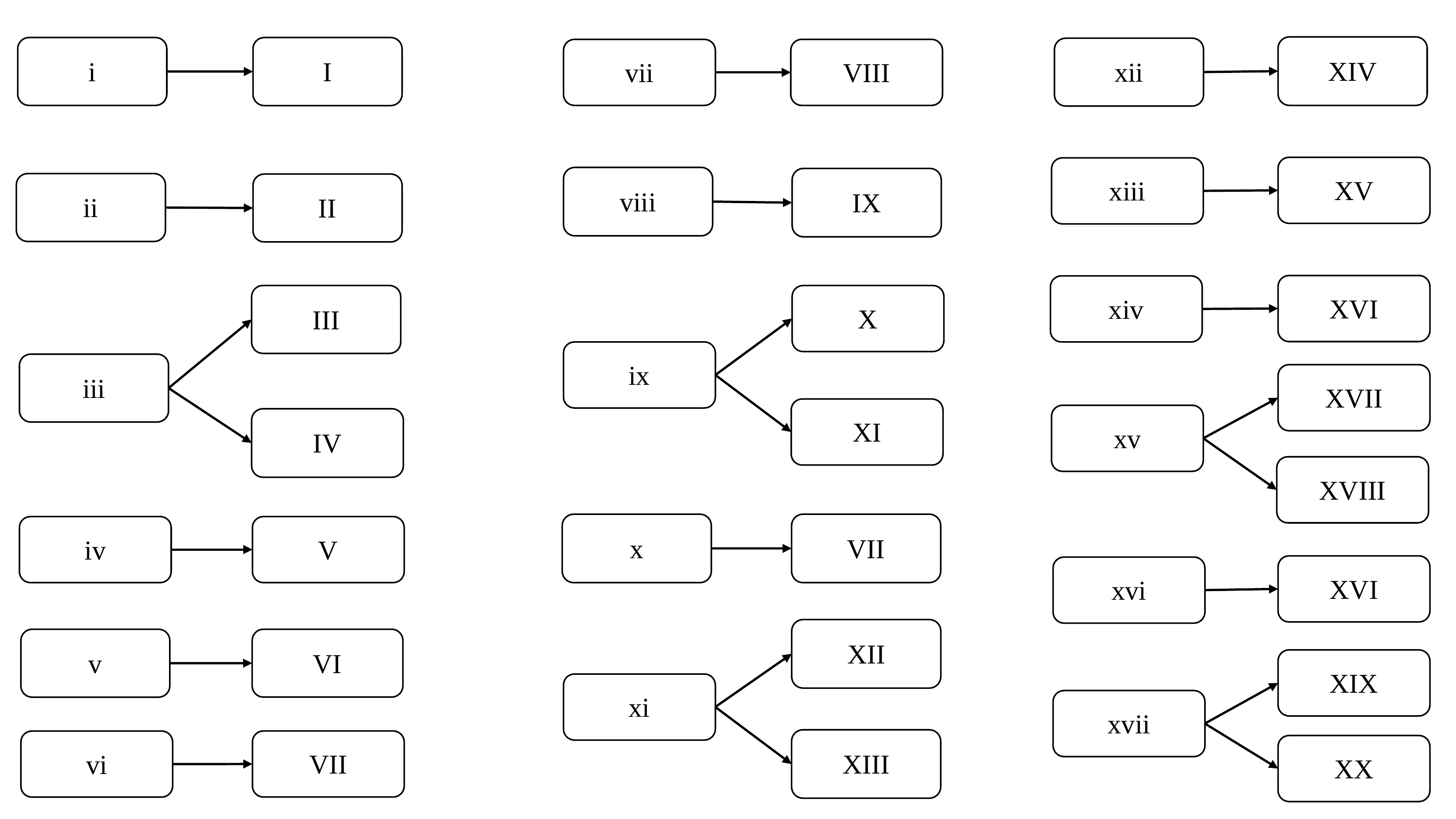}}
\caption{Extensions of semidegenerate systems} \label{extension}
\end{figure}

Figures \ref{4thorder} and \ref{extension} depict the B\^ocher contraction scheme for 4th order extensions of semidegenerate systems.
\section{Conclusions and outlook}\label{conclusions}
Using the powerful tool of B\^ocher contractions we have found a family of 20 semidegenerate 2nd order 3D conformally
superintegrable Laplace 
systems and 17 4th order conformally superintegrable Laplace systems that are extensions of these and have related them via B\^ocher contractions.
These correspond to about 100 Helmholtz systems on a variety of manifolds. These results apply to classical systems with only a few obvious adjustments.
Every semidegenerate system extends to a 4th order system. Only the B\^ocher contraction $[11111]\downarrow [5]$ fails to produce any new semidegenerate system.
This work partially fills a gap in the classification of 3D 2nd order 
superintegrable systems. 

The difficulty here is that, as yet, there is no detailed structure and classification theory for semidegenerate systems 
or for 4th order superintegrable systems. For nondegenerate 3D systems there is a complete theory with a guarantee that any B\^ocher contraction 
of a nondegenerate system yields a nondegenerate system, unless the contracted potential is functionally dependent. Here we can use 
B\^ocher contractions as a valuable  calculational
tool but with no guarantee of completeness. Is every semidegenerate system obtainable from (\ref{Hcoulsphere}) by a sequence
of B\^ocher contractions and St\"ackel transforms? We suspect so but have no proof. Does every semidegenerate system extend to a 4th 
order superintegrable system? Again, we suspect so but have no proof. System ii and all systems obtained from it by contraction have a closed symmetry algebra. Is the symmetry algebra of i closed? We expect so but have not carried out the difficult calculation to verify this.

\section{Acknowledgment}
This work was partially supported by a grant from the Simons Foundation (\#412351, Willard Miller, Jr)  and by CONACYT grant (\# 250881 to M.A. Escobar ).

\end{document}